\documentclass[aps,prb,reprint,groupedaddress]{revtex4-2}

\usepackage{dcolumn}
\usepackage{bm}
\usepackage[section]{placeins}
\usepackage{color}
\usepackage{mathrsfs}
\usepackage{ulem}
\usepackage{hyperref}
\usepackage{graphicx}
\usepackage{svg}
\usepackage{braket}
\usepackage{here}
\usepackage{placeins}
\usepackage{float}

\newcommand{\tb}[1]{\textcolor{black}{#1}}

\begin{document}


\title{Coherent control of solid-state defect spins via patterned boron-doped diamond circuit}


\author{Masahiro Ohkuma$^{1}$}
\thanks{These authors contributed equally to this work.}
\author{Eikichi Kimura$^{1}$}
\thanks{These authors contributed equally to this work.}
\author{Eunsang Lee$^{2}$}
\thanks{These authors contributed equally to this work.}
\author{Ryo Matsumoto$^{3}$}
\email{MATSUMOTO.Ryo@nims.go.jp}
\author{Shumpei Ohyama$^{1}$}
\author{Saki Tsuchiya$^{1}$}
\author{Harim Lim$^{2}$}
\author{Yong Soo Lee$^{2}$}
\author{Yoshihiko Takano$^{3}$}
\author{Junghyun Lee$^{2}$}
\email{jh\_lee@kist.re.kr}
\author{Keigo Arai$^{1}$}
\email{arai.k.835f@m.isct.ac.jp}
\affiliation{$^{1}$School of Engineering, Institute of Science Tokyo, Yokohama, Kanagawa 226-8501, Japan}
\affiliation{$^{2}$Center for Quantum Technology, Korea Institute of Science and Technology, Seoul, 02792, Republic of Korea}
\affiliation{$^{3}$Research Center for Materials Nanoarchitectonics (MANA),National Institute for Materials Science, Tsukuba, Ibaraki 305-0047, Japan}

\date{\today}

\begin{abstract}
Monolithic integration, which refers to the incorporation of all device functionalities within a single material, shows significant potential for creating scalable solid-state quantum devices.
This study demonstrated the coherent control of nitrogen-vacancy (NV) spins using an electronic circuit monolithically integrated within diamond: a patterned, conductive boron-doped diamond (BDD) microwave waveguide.
First, we validated the high-frequency performance of the circuit by characterizing its impedance up to the microwave range, confirming its capability for efficient microwave transmission.
Then, using this monolithically integrated BDD--NV hybrid system, we performed optically detected magnetic resonance and observed noticeable Rabi oscillations driven by the metallic BDD circuit.
Importantly, we verified that the BDD antenna has a minimal detrimental impact on the NV spins; microwave-induced heating is negligible under both pulsed and continuous driving, and the spin relaxation time ($T_1$) remains unperturbed.
This approach paves the way for a new class of compact, robust, and versatile quantum platforms suitable for sensing and information processing in various environments.
\end{abstract}

\maketitle

\section{Introduction}
Diamond is a versatile material capable of accommodating various impurities that substantially alter its electronic, thermal, and optical properties.
Among these, boron doping is a well-known technique for transforming diamond into a p-type semiconductor.
With sufficient doping, diamond can exhibit metallic behavior and superconductivity at cryogenic temperatures \cite{ekimovSuperconductivityDiamond2004, takanoSuperconductivityDiamondThin2004, yokoyaOriginMetallicProperties2005, takanoSuperconductivityCVDDiamond2009}.
Boron-doped diamond (BDD) can be grown into thin films via microwave plasma chemical vapor deposition, enabling the development of advanced devices, such as superconducting quantum interference devices and microwave components \cite{mandalDiamondSuperconductingQuantum2011, kageuraSinglecrystallineBorondopedDiamond2019, oripovLargeMicrowaveInductance2021, cuencaSuperconductingBoronDoped2023}.
These BDD thin films support the transmission of alternating electric currents across a broad frequency range, thereby showing great potential for use in electric circuits.
With such unique electrical and physical properties, BDD holds great promise for cutting-edge applications in both power electronics \cite{isbergHighCarrierMobility2002, geisProgressDiamondPower2018, donatoDiamondPowerDevices2019} and quantum technologies \cite{dasDiamondUltimateMaterial2022}.

The negatively charged nitrogen-vacancy (NV) center---a point defect in the diamond lattice---has become a leading platform for quantum measurement, owing to its unique properties \cite{jelezkoObservationCoherentOscillations2004, gaebelRoomtemperatureCoherentCoupling2006, jelezkoObservationCoherentOscillation2004, hansonPolarizationReadoutCoupled2006, duttQuantumRegisterBased2007, taylorHighsensitivityDiamondMagnetometer2008a, degenScanningMagneticField2008, balasubramanianNanoscaleImagingMagnetometry2008a, mazeNanoscaleMagneticSensing2008, acostaDiamondsHighDensity2009}.
One of its most striking features is the ability to maintain coherent spin control under extreme conditions: from ultrahigh vacuum to pressures as high as 130 GPa \cite{schaefer-nolteDiamondbasedScanningProbe2014, bhattacharyyaImagingMeissnerEffect2024, wangImagingMagneticTransition2024} and temperatures ranging from cryogenic to a blazing 1400~K \cite{scheideggerScanningNitrogenvacancyMagnetometry2022a, liuCoherentQuantumControl2019, kubotaWideTemperatureOperation2023, fanQuantumCoherenceControl2024}.
Additionally, NV control systems can be miniaturized, integrated, and made portable, rendering it suitable in various practical applications \cite{kimCMOSintegratedQuantumSensor2019, sturnerIntegratedPortableMagnetometer2021, omirzakhovIntegratedReconfigurableSpin2023, kumarHighDynamicrangePortable2024}.
These applications rely on the use of alternating electromagnetic fields across a broad frequency spectrum: gigahertz microwaves for manipulating electronic spin ground states \cite{dohertyNitrogenvacancyColourCentre2013}, megahertz radio frequencies for nuclear spin control \cite{jiangRepetitiveReadoutSingle2009, neumannSingleShotReadoutSingle2010, lovchinskyNuclearMagneticResonance2016}, and kilohertz currents for modulating electronic spin states \cite{araiFourierMagneticImaging2015, zhangSelectiveAddressingSolidstate2017}.
These fields are generated \tb{using metal-based circuits made from materials such as gold or copper, often layered atop a thin titanium adhesive film}.
However, metal circuits have several limitations; they are prone to scratching, susceptible to corrosion, and require chemical etching to remove the underlying adhesive.
therefore, replacing them with conductive BDD in these circuits can significantly improve their durability while simplifying the overall system structure.

In this study, we \tb{used an $\Omega$-shaped BDD thin film } as a source of alternating magnetic fields to coherently manipulate an ensemble of NV centers in diamond.
Our \tb{key} findings \tb{are as follows}: (i) the impedance of the BDD thin films was \tb{accurately} modeled \tb{by a lumped parameter circuit model up to 3 GHz;} (ii) \tb{BDD in the metallic state} with high boron concentration \tb{generated} microwave fields \tb{comparable in strength to} those produced by conventional metal wires, and (iii) heating effects \tb{remained minimal, even under} both \tb{short microwave pulses} and long continuous microwave drives.
\tb{These results show that BDD offers a robust and scalable platform for NV center control, particularly under extreme conditions.}
\tb{Furthermore, its compatibility with homoepitaxial diamond growth renders it suitable for seamless on-chip integration, improved thermal management, and simplified fabrication workflows.}
In the following sections, we provide details of our experimental results and discuss the implications \tb{of this technology}.

\section{Overview of BDD--NV hybrid system}
\begin{figure}[htb!]
\centering\includegraphics[clip,width=1\columnwidth]{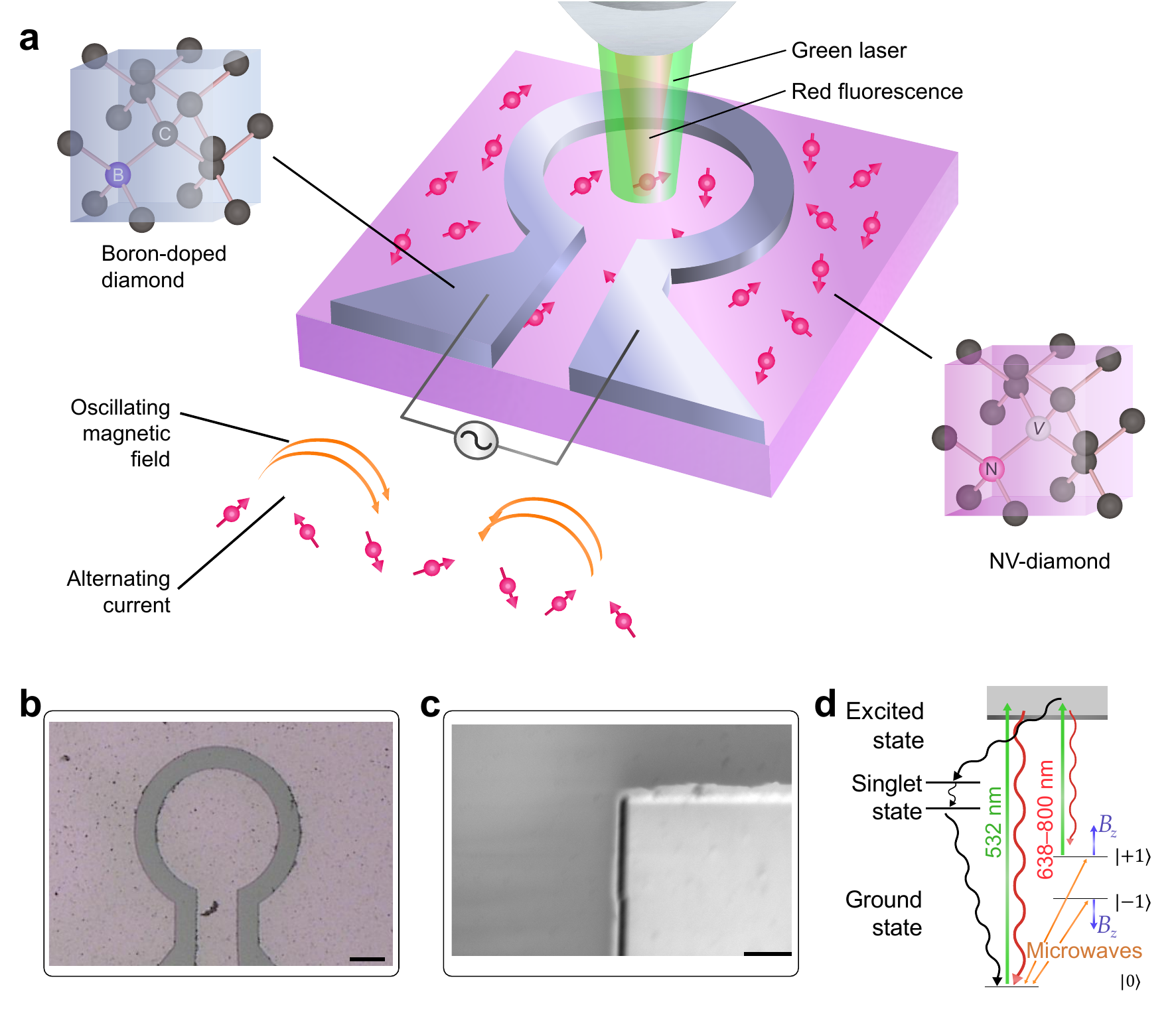}
\caption{\label{f1} (a)~(Top) Overview of BDD--NV system comprising a $\Omega$-shaped BDD circuit fabricated on a diamond substrate containing an ensemble of NV centers.
In this study, we measured NV centers mainly inside the $\Omega$ area.
Crystal structure of BDD is shown in the left side.
Structure of NV center in diamond lattice is shown in the right side.
N and $V$ denote nitrogen atom and vacancy, respectively.
(Bottom) Side view of BDD--NV system.
A BDD thin film is grown on a single-crystalline diamond containing NV centers.
The $\Omega$-shaped BDD circuit hosts an alternating current at the microwave frequency range, generating an oscillating magnetic field.
This microwave field is used for manipulating the NV centers in the diamond substrate.
(b)~Optical micrograph of BDD circuit with boron concentration of $3\times10^{21}$ $\rm cm^{-3}$.
Scale bar is 50 $\mu$m.
(c)~Scanning electron microscope image of an edge of BDD thin film grown on single-crystalline diamond substrate with NV centers.
Scale bar is 10 $\mu$m.
(d)~Energy diagram of NV center.
The ground states can be excited optically by a 532~nm laser and decay along with spin-dependent fluorescence.
The spin state is manipulated by a microwave magnetic field applied by a BDD circuit.}
\end{figure}

The BDD--NV hybrid system comprises an $\Omega$-shaped BDD circuit fabricated on a diamond substrate containing an ensemble of NV centers, as illustrated in Fig.~\ref{f1}.
This study examines three variants of boron-doping densities, labeled BDD1, BDD2, and BDD3, corresponding to decreasing boron concentrations from metal to insulator, as presented in Table~\ref{t1}.
The BDD circuits were produced in an $\Omega$ configuration with an inner diameter of 140~$\mu$m and thickness of 700~nm via masked microwave plasma chemical vapor deposition \cite{takanoSuperconductivityDiamondThin2004, matsumotoNoteNovelDiamond2016}.
The boron concentrations were determined via an analysis of the temperature dependence of resistivity.

The diamond substrate measures 3~mm $\times$ 3~mm $\times$ 0.5~mm and features a concentration of NV centers at 4.5~ppm across all four crystallographic axes.
The NV center possesses an electronic ground-state triplet characterized by the states $\ket{0}$ and $\ket{\pm1}$, with a separation of 2.87~GHz resulting from spin--spin interactions.
Additionally, the $\ket{\pm1}$ states experience further splitting due to the Zeeman effect when subjected to a finite magnetic field $B$ aligned with the NV axis of interest.
The energy levels corresponding to these spin states can be analyzed via optically detected magnetic resonance (ODMR) spectroscopy, which involves observing changes in the red fluorescence of the NV center as the microwave field frequency matches the energy difference between the $\ket{0}$ state and either of the $\ket{\pm1}$ states \cite{oortOpticallyDetectedSpin1988, gruberScanningConfocalOptical1997}.

The $\Omega$-shaped BDD circuit serves as a versatile antenna, designed to generate oscillating magnetic fields across a broad spectrum of frequencies required for quantum control.
This wideband operation is experimentally confirmed via direct current transport measurements and the coherent manipulation of NV center spins at microwave frequencies.
For these experiments, the BDD circuit connects to electrodes on a printed circuit board via a $\o=25$ $\mu$m gold wire and silver paste.
The printed circuit board is linked to a signal generator through a Sub Miniature A cable, ensuring a matched impedance of 50~$\Omega$.
The microwave field produced by this BDD circuit enables the manipulation of the triplet ground states of the NV centers, with this study concentrating on the NV centers situated within the $\Omega$-shaped area.

\begin{table}[htb!]
\caption{\label{t1} Parameters of BDD circuits. $\rho$ and R.~T.~denote resistivity and room temperature, respectively. Boron concentration is estimated based on the temperature dependence of resistivity.}
\begin{ruledtabular}
\begin{tabular}{cccc}
 & BDD1& BDD2& BDD3 \\
\hline
$\rho$ ($\Omega$ cm) at R. T.& $9.8\times10^{-4}$ & $4.1\times10^{-3}$ & $5.8\times10^{-1}$\\
B concentration ($\rm cm^{-3}$) & $3\times10^{21}$ & $4\times10^{20}$& $1\times10^{20}$\\
Thickness (nm) & 780 & 640 & 690 \\
Conducting state & Metal & Metal& Insulator\\
\end{tabular}
\end{ruledtabular}
\end{table}

\section{Evaluation of electrical properties of BDD}
\begin{figure}[htb!]
\centering\includegraphics[]{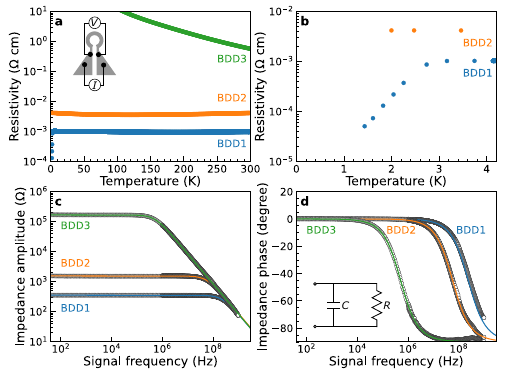}
\caption{\label{f_RT} (a) Temperature dependence of electrical resistivity of BDD for three different boron concentrations. The BDDs have resistivities of $9.8\times10^{-4}$, $4.1\times10^{-3}$, and $5.8\times10^{-1}$ $\Omega$ cm at room temperature, respectively. BDD1 and BDD2 exhibit metallic behaviors, whereas BDD3 exhibits an insulating behavior.
The inset shows the four-terminal arrangement used for the measurement method. 
(b) Temperature dependence of the electrical resistivity of BDD1 and BDD2 below 4.2 K. A drop in the resistivity of BDD1 indicates a superconducting transition.
(c) and (d) Frequency dependence of impedance (c)~amplitude and (d)~phase on BDDs.
Each data point is obtained using a coaxial probe directly touching the BDD circuits.
Notably, the values from 40~Hz to 110~MHz and 1~MHz to 1~GHz were obtained using different measurements (refer to the methods section for specifics).
Solid lines represent fitted curves based on the model of parallel resistance and capacitance, as shown in inset of (d).}
\end{figure}

To evaluate the boron concentration of BDD, we first performed electrical resistivity measurements on the BDD circuits, as illustrated in Figs.~\ref{f_RT}a and b.
The electrical resistivity was assessed as a function of temperature using the four-probe technique enclosed in the Physical Properties Measurement System or a home-built Gifford--McMahon cryostat system.
BDD1 exhibited metallic characteristics near room temperature, with resistivity decreasing as the temperature decreased.
This trend aligns with findings from existing studies \cite{ekimovSuperconductivityDiamond2004, takanoSuperconductivityCVDDiamond2009}.
Notably, at approximately 3~K, BDD1 exhibited a resistivity drop indicative of a superconducting transition.
BDD2 exhibited a similar pattern but without any superconducting transition above 2~K, whereas BDD3, with reduced boron concentration, exhibited typical insulating behavior, characterized by an increase in resistivity with decreasing temperature.
Studies have reported an insulator-to-metal transition at a boron concentration of approximately $3\times10^{20}~\rm {cm}^{-1}$ \cite{yokoyaOriginMetallicProperties2005, takanoSuperconductivityCVDDiamond2009}.
According to the studies \cite{takanoSuperconductivityCVDDiamond2009, zhangCharacterizationHeavilyBorondoped1996, lagrangeActivationEnergyLow1998}, the boron concentrations were estimated to be $3\times10^{21}$, $4\times10^{20}$, and $1\times10^{20}$ $\rm {cm}^{-1}$, categorizing BDD1 and BDD2 as metals and BDD3 as an insulator, respectively.

Next, we assessed the impedance of the BDD circuits across a frequency range of 40~Hz to 1~GHz.
The frequency-dependent variations in impedance amplitude and phase for BDDs, illustrated in Figs.~\ref{f_RT}c and d, were recorded from 40~Hz to 110~MHz and from 1~MHz to 1~GHz using distinct equipment (refer to the method section for specifics).
Consequently, the data points between 1~MHz and 110~MHz exhibit overlap.
Impedance measurements were performed using a coaxial probe that made direct contact with the BDD circuit, with each data point representing the average of five measurements.
Both the amplitude and phase of the impedance decline in the high-frequency range, with the onset of this decrease being contingent upon the resistivity of the BDD circuits.

This phenomenon indicates a straightforward lumped parameter circuit model comprising a parallel resistance and capacitance, as illustrated in the inset of Fig.~\ref{f_RT}d.
This lumped parameter circuit model has a valid design because the $\sim$200~$\mu$m dimension of the BDD circuit is significantly smaller than the $\sim$10~cm wavelength of 3~GHz microwave fields.
The parasitic capacitance component $C$ is connected in parallel to the resistance component $R$ of the BDD circuit, as depicted in the inset of Fig.~\ref{f_RT}d, enabling the frequency $f$ dependence of the impedance $Z$ and the phase $\theta$ to be expressed as $Z(f)=(1/R+2j\pi fC)^{-1}$ and $\theta(f) = \tan^{-1}(2\pi fRC)$, respectively, which behaves as a high-pass filter.
The measured data fit well with the proposed model, with the fitted values presented in table~\ref{t_z}.
In particular, the obtained $R$ values are significantly consistent with those obtained from resistivity measurements using the four-probe method.
$C$ is expected to vary with the circuit thickness, which correlates with the electrode size, given that the length and width of the circuit are approximately equal.
Notably, the fitted value of $C$ increases with increasing circuit thickness.
\begin{table}[htb!]
\caption{\label{t_z} Fitted values of impedance measurements. All error values reported in this paper are estimated based on 1.96$\sigma$ of parameter fitting, where $\sigma$ is the standard deviation.}
\begin{ruledtabular}
\begin{tabular}{cccc}
 & $R$ ($\Omega$)& $C$ (pF)\\
\hline
BDD1& $(3.547 \pm 0.001) \times 10^2$ & $1.890 \pm 0.004$\\
BDD2 & $(1.552 \pm 0.002) \times 10^3$ & $1.778 \pm 0.008$\\
BDD3 & $(1.6680 \pm 0.0005) \times 10^5$ & $1.854 \pm 0.002$\\
\end{tabular}
\end{ruledtabular}
\end{table}

\section{Manipulation of NV center using BDD circuits}
Impedance measurements guided the design of a microwave circuit for driving NV centers, comprising an $\Omega$-shaped BDD structure on a diamond substrate hosting an NV ensemble. The circuit operates without impedance matching, offering design flexibility for integration and deployment in extreme environments. At 2.87 GHz, the BDD1 circuit exhibits an impedance of 28.6~$\Omega$ and resistance of 355~$\Omega$, whereas the capacitor $C$ exhibits an impedance of 28.7~$\Omega$. These elements form a parallel circuit, enabling partial microwave transmission through the resistive path to control the NV centers. The irradiated microwave power is modulated by the resistance, which scales with the boron doping level.

\begin{figure}[htb!]
\centering\includegraphics[]{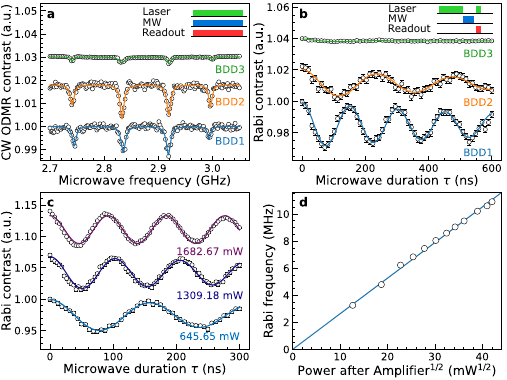}
\caption{\label{f_ODMR} (a) ODMR spectra measurement result using different BDDs with varied boron densities. The spectra are offset by 0.015 for BDD2 and 0.03 for BDD3. (b) Rabi oscillations using different BDD circuits and the pulse sequence used for Rabi measurement (inset). Rabi contrasts are offset by 0.02 for BDD2 and 0.04 for BDD3. These signals are obtained at the center of each $\Omega$-shaped circuit. Each spectrum is normalized by its respective maximum intensity. (c) Rabi oscillations measured in BDD1 antenna at different input powers: 517.61, 1309.18, and 1682.67 mW. The data are vertically offset by 0.07 for $P = 1309.18$ mW and 0.14 for $P = 1682.67$ mW for clarity. The corresponding Rabi frequencies at each input power were 6.83, 9.48, and 10.6 MHz, respectively.
(d) Input microwave power dependence of Rabi oscillation frequency. The result shows $\Omega \propto \sqrt{P_{\rm MW}}$.}
\end{figure}

First, continuous-wave (CW) ODMR and Rabi oscillation measurements between the $\ket{0}$ and $\ket{-1}$ states are performed using BDD-based $\Omega$-shaped circuits under a static magnetic field of ~4.5~mT. Figs.~\ref{f_ODMR}a and b show data collected near the centers of three BDD structures. The measured ODMR contrasts and FWHMs at $\sim$2.7~GHz are ($0.8\pm0.2$)\% and $4\pm1$~MHz (BDD1), ($0.9\pm0.1$)\% and $4\pm1$~MHz (BDD2), and ($0.20\pm0.09$)\% and $2\pm1$~MHz (BDD3), respectively. Rabi frequencies were $6.62\pm0.04$~MHz for BDD1 and $4.19\pm0.06$~MHz for BDD2 at a microwave input power of 46.5 dBm, consistent with the impedance data indicating enhanced microwave delivery in lower-resistivity samples. No Rabi oscillation is observed in BDD3 owing to its high resistivity.
Moreover, Rabi oscillations in BDD1 were measured at input powers of 517.61, 1309.18, and 1682.67 mW, as shown in Fig.~\ref{f_ODMR}c.
The corresponding Rabi frequencies are 6.83, 9.48, and 10.6 MHz, respectively.
These data are plotted in Fig.~\ref{f_ODMR}d, which shows the expected linear relationship between the square root of the applied microwave power and Rabi frequency.

To assess the suppressed thermal effects of the BDD microwave antenna during NV spin control, we performed two measurements: long-duration Rabi chevron mapping and off-resonant microwave driving followed by Ramsey interferometry. Previous studies have shown that temperature variations in the sample predominantly shift the zero-phonon line (ZPL) of the NV center, affecting the spin states transition resonance frequencies \cite{acostaTemperatureDependenceNitrogenVacancy2010}.
Heating-induced frequency shifts from metallic microwave antennas have been previously reported \cite{wangMicrowaveHeatingEffect2022}. Efficient heat dissipation is expected at the BDD--NV hybrid diamond interface, owing to its monolithic structure.

\begin{figure}[htb!]
\centering\includegraphics[]{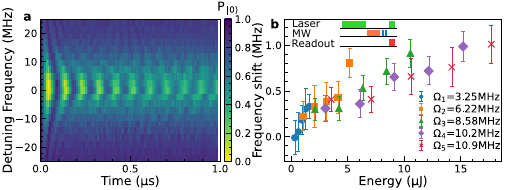}
\caption{\label{f_rabi} (a) Rabi Chevron measurement result using BDD1 antenna. Rabi oscillations are measured by sweeping the microwave frequency from $-25$ to $+25$ MHz while fixing the microwave duration at 1 $\mu$s. $\rm P_{\ket{0}}$ represents the probability of preserving the initial state $\ket{0}$.
(b) Plot of frequency shift as a function of the total applied energy, along with the pulse sequence used for the measurement (inset).
To quantify the heating effect, the $\pi$/2 pulses in the Ramsey sequence are detuned by $+3$ MHz from the NV resonance frequency (blue box in the inset), and an additional microwave pulse detuned by 400 MHz was applied before the Ramsey sequence (orange box in the inset).
The total energy, shown on the horizontal axis, is calculated as $E = P \times t$, where $P$ is the applied microwave power, and is the pulse duration (i.e., the width of the orange box).
The legends indicate the Rabi frequencies corresponding to each input power.
A longer heating time results in a larger frequency shift.}
\end{figure}

To probe potential resonance shifts during rapid driving, we performed a full Rabi chevron measurement (Fig.~\ref{f_rabi}a), sweeping the microwave frequency from $–25$ to $+25$~MHz around the resonance at 2.7405~GHz with fixed microwave signal duration 1~$\mu$s. The resulting chevron exhibits a symmetric pattern, in contrast to simulation data where the thermal perturbation asymmetries the chevron pattern (see Supplemental Information). These results confirm that, within short-duration (1~$\mu$s) Rabi experiments, the BDD microwave antenna induces negligible heating impact on the NV centers.

To quantify heating effects from prolonged microwave exposure, we then applied an off-resonant microwave pulse between the initialization and Ramsey sequences (inset in Fig.~\ref{f_rabi}b).
The total microwave energy $E = P \times t$ was varied by tuning both power (via expected Rabi frequency) and duration, and the resulting Ramsey frequency shifts were extracted using a fast Fourier transform (FFT) analysis.
As the total input energy increased, the Ramsey FFT peak shifted to higher frequencies, capturing the negative shift in ZPL with temperature rise (Fig.~\ref{f_rabi}b).
Notably, the maximum frequency shift observed for a 10~$\mu$s microwave pulse duration with an 11~MHz Rabi frequency is approximately 1~MHz (See Supplemental Information for the more details).

The observed frequency shift ($\sim$1~MHz) is smaller than the hyperfine splitting induced by NV host nuclear spins, such as $^{14}$N (2.173~MHz) or $^{15}$N (3.3~MHz), and is significantly less than the maximum applied Rabi frequency of 11~MHz.
Thus, the heating effect from the BDD-based microwave delivery can be considered negligible in experiments involving continuous microwave driving, such as Floquet engineering \cite{boyersFloquetengineeredQuantumState2019, nishimuraFloquetEngineeringUsing2022} or spin-locking for dynamic nuclear polarization \cite{rizzatoPolarizationTransferOptically2022, huExperimentalRealizationDeepsubwavelength2018}.

\begin{figure}[htb!]
\centering\includegraphics[]{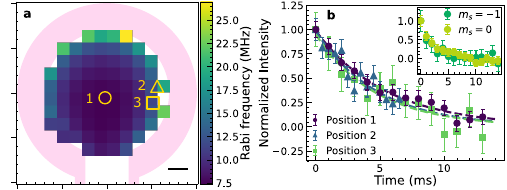}
\caption{\label{f5} (a) Spatial distribution of the measured Rabi frequency inside $\Omega$-shaped area.
The orange marks indicate the measurement positions corresponding to those shown in (b), \tb{where position 3 is as close as 5~$\mu$m away from the BDD antenna}.
Black scale bar is 20 $\mu$m.
(b) $T_1$ measurement at three positions: the center (purple circle) and the two edges (deep blue triangle and light green rectangle) of the loop.
The inset shows spin relaxation curves for the $\ket{0}$ (light olive green circle) and $\ket{-1}$ (green circle) states at position 3.
The intensity values on the vertical axis were calculated by normalizing the difference between the two spin relaxation data of the $\ket{0}$ and $\ket{-1}$ states.}
\end{figure}

Finally, to assess the spatial uniformity of the microwave field, Rabi oscillations were mapped across the $\Omega$-shaped BDD1 region. As shown in Fig.~\ref{f5}a, the interior was divided into a $10\times10$ pixel grid, revealing detectable oscillations throughout.
The Rabi frequency exhibited a minimum at the center ($\sim$10 MHz), increasing toward the edges ($\sim$20 MHz), which is consistent with simulations presented in Supplementary Section S2.

The observed fast Rabi oscillations confirm the suitability of NV centers near the BDD antenna for measurements requiring rapid spin control.
To evaluate the impact of microwave-induced magnetic Johnson noise from fluctuating currents, $T_1$ relaxation times are measured at multiple positions within the BDD1 circuit (orange markers in Fig.~\ref{f5}a), including both the loop center and \tb{the edge as close as 5~$\mu$m}.
Fig.~\ref{f5}b shows $T_1$ decay curves with consistent relaxation rates across positions, within experimental uncertainty.
Additionally, comparable $T_1$ values are observed for both the $m_s = 0$ and $m_s = -1$ spin states (inset, Fig.~\ref{f5}b).
Previous reports indicate that proximity to metallic structures reduces $T_1$ and introduces state-dependent relaxation asymmetries \cite{kolkowitzProbingJohnsonNoise2015}.
By contrast, the measured $T_1$ values indicate that BDD-based microwave circuits minimally perturb NV spin relaxation, indicating negligible magnetic Johnson noise, and enabling fast, coherent control close to the BDD antenna.

\section{Discussion}
\tb{In this study, we demonstrated that coherent control of NV centers can be achieved using a BDD-based circuit, without precise impedance matching, demonstrating the inherent robustness of this approach.
The impedance behavior observed in our system is well described by a parallel $R$--$C$ circuit model, offering a solid foundation for future improvements.}
For instance, further optimization of the impedance matching, achievable by tuning the circuit geometry to adjust both $R$ and $C$, \tb{could lead to substantial performance gains, such as increased Rabi frequencies or reduced power consumption.}
\tb{Another key route in enhancing circuit efficiency is minimizing the resistivity of the BDD material because lower resistivity improves current flow through the resistive channel $R$, consequently strengthening the generated oscillating magnetic fields.}
\tb{Previous studies have shown} that boron concentration can \tb{reach up} to $1\times10^{22} \rm{cm}^{-3}$ \cite{kawanoSuperconductortoinsulatorTransitionBorondoped2010}, \tb{pushing the Hall carrier concentration to similar levels.} 
\tb{Achieving such high doping densities shows strong potential for further reducing the resistivity of BDD, rendering it even more effective for high-performance quantum control applications.}

\tb{Employing BDD circuits for microwave delivery offers several key advantages, particularly their exceptional physical and chemical robustness, as well as their compatibility with integrated device architectures.}
\tb{BDD is chemically stable across a wide pH range, rendering it ideal for use in both acidic and alkaline environments; an asset that has already proven valuable in electrochemical applications} \cite{yanoElectrochemicalBehaviorHighly1998, gandiniOxidationCarboxylicAcids2000}.
Moreover, it \tb{can withstand extreme conditions, including} high temperatures and pressures, and remains reusable after returning to ambient conditions \cite{matsumotoNoteNovelDiamond2016, matsumotoElectricalTransportMeasurements2020, matsumotoEmergenceSuperconductivity202024}.
Furthermore, BDD circuits can be precisely microfabricated.
\tb{By introducing insulating layers of undoped diamond between BDD elements, these circuits can also be vertically stacked by depositing insulating undoped diamond between them, enabling highly compact and integrated device designs.}
This enables the development of compact and resilient quantum sensors that \tb{combine} NV centers with \tb{BDD circuitry}, presenting a promising avenue for quantum sensing in challenging environments such as \tb{industrial sites, deep-sea exploration, and outer space} \cite{fuSensitiveMagnetometryChallenging2020a}.

\tb{Our electrical measurements indicate that BDD circuits are highly effective at generating microwave fields and producing oscillating electromagnetic fields at lower frequencies, broadening their applicability across multiple domains.}
For instance, \tb{kilohertz-range alternating currents are employed} in phase-encoding techniques \cite{araiFourierMagneticImaging2015, zhangSuppressionPhotonShot2017}.
\tb{Meanwhile,} megahertz \tb{radio-frequency} waves \tb{are used to manipulate} different nuclear spins both \tb{within diamond and in surrounding materials} \cite{jiangRepetitiveReadoutSingle2009, neumannSingleShotReadoutSingle2010, lovchinskyNuclearMagneticResonance2016, bruckmaierImagingLocalDiffusion2023}.
Owing to the unique thermal conductivity of BDD \cite{prikhodkoLowTemperatureThermal2019}, the heat produced by these alternating currents can be efficiently dissipated \tb{under ambient conditions}.
\tb{This thermal stability, combined with the} extensive range of operational frequencies, further enhances the appeal of BDD circuits, indicating promising prospects in quantum sensing and quantum information processing.

\begin{acknowledgments}
The authors would like to thank H. Sano for helping with impedance measurements. This work is supported by, or in part by, JSPS KAKENHI Grant numbers JP21K14524 and JP23K26528, JSPS Fostering Joint International Research Grant number JP23KK0267, JSPS Bilateral Program Grant number JPJSBP120238803, JST K Program Grant number JPMJKP24F3, the National Research Foundation (NRF) of Korea program under Grant No. RS-2024-00442710, and KIST institutional program (Project. No. 2E33541).
M.O. receives funding from JSPS Grant-in-Aid for JSPS Fellows Grant number JP24KJ1035.
E.K. receives funding from JST SPRING, Japan Grant Number JPMJSP2180.
\end{acknowledgments}
\section*{Author contributions}
M.O., R.M., and K.A. designed the study. M.O., E.K., E.L., R.M. S.O., S.T., and H.L. conducted electrical, optical, AFM, and SEM measurements, analyzed data, and conducted numerical simulations. R.M. fabricated the BDD device. J.L., Y.T., and K.A. conceived the application of the BDD to ODMR and supervised the project. All authors discussed the results and participated in writing the manuscript. 
All the authors discussed the results and commented on the manuscript.
\section*{Competing interests}
The authors have no competing interests to disclose.

\section{Method}
\subsection{BDD-NV sample fabrication}
A [100]-oriented single-crystalline diamond substrate purchased from Element Six (DNV-B14) was used in this study.
BDD was deposited on the diamond substrate using microwave plasma chemical vapor deposition, the procedure of which is described elsewhere \cite{takanoSuperconductivityDiamondThin2004}.
We repeated deposition processes three times to prepare three $\Omega$-shaped BDDs with different boron concentrations on the same diamond substrate.
The BDD circuits were fabricated via the following steps: (1) Before deposition of the metal mask, it is treated with hot mixed acid to oxygenate the termination. The diamond substrate is treated with a hot mixture of sulfuric acid: nitric acid = 1:3. (2) Electron beam resist is coated. (3) Circuit pattern is fabricated by electron beam lithography. (4) The metal mask is prepared via deposition of a metal mask Ti/Au thin film and lift-off method and then annealed under vacuum ($10^{-5}$ Torr) at 450 ${}^\circ$C for 30 min to form a TiC adhesive layer at the interface between the metal mask and diamond. (5) Deposition of the BDD circuit by microwave plasma chemical vapor deposition. (6) Removal of the metal mask by hot mixed acid treatment. We repeated this process three times to prepare three $\Omega$-shaped BDDs with different boron concentrations on the same diamond substrate. The boron concentration was controlled by changing the ratio of ﬂow rate in methane as a carbon source and trimethylboron as a boron source. The inner and outer diameters were 140 $\mu$m and 200 $\mu$m, respectively. All BDDs had the same shape.
The thickness of the BDD thin film was measured using an atomic force microscope (Nanocute, SII NanoTechnology Inc.) at room temperature.
Scanning electron microscope images were obtained using JSM-6010LA, JEOL, with an acceleration voltage of 5 keV.

\subsection{BDD electrical measurements}
The temperature dependence of electrical transport was measured using the current-driven four-probe method with a home-built GM cryostat system for BDD1 and Physical Properties Measurement System (Quantum Design) for BDD2 and BDD3. In the home-built GM cryostat system, an ADCMT 6146 current source and a KEITHLEY 2000 voltage meter were used for measurements.
First, the temperature was decreased from 300 to 4.2~K, and the chamber pressure was decreased using a scroll-type vacuum pump to reach temperatures below 2~K.

The frequency dependence of the impedance was measured using two impedance analyzers (4294A and E4991A, Agilent) at room temperature.
We used 4294A from 40~Hz to 110~MHz and E4991A from 1~MHz to 1~GHz.
The coaxial probe (CPHF-R-0.5, Yokowo DS) connected to the impedance analyzer was directly put in contact with the BDD circuits to measure the impedance.
Each measurement was repeated five times, and their average values were used to minimize the slight differences between different samples.

\subsection{BDD-NV spin measurements}
CW ODMR and Rabi oscillation measurements using the BDD microwave antenna were performed using a home-built confocal microscope.
A 532-nm green laser (MGL-III-532-200mW, CNI Laser) was pulsed using an acousto-optic modulator (AOM, AOMO 3200-121, Gooch$\&$Housego) and irradiated the diamond through a 0.95 NA air objective lens (EC Epiplan-Apochromat 100x/0.95 HD DIC M27, Zeiss).
The emitted red fluorescence was collected using the same objective lens and detected by a single-photon counting module (SPCM-780-33-BR1, Excelitas).
A data acquisition device (NI 6363 and NI PXIe-1073, National Instruments) was used for fluorescence counting.
A vector signal generator (SMW 200A, Rohde $\&$ Schwarz) generated microwave signals for CW measurements, and an arbitrary waveform generator (AWG70002A, Tektronix) generated signals for pulsed measurements.
These microwave signals were selectively passed through a switch (ZASWA-2-50DR$+$, MiniCircuit) and amplified using an amplifier (ZHL16W-43-S$+$, Mini-Circuit).
For pulsed measurements, a field programmable gate array is used to gate the AOM switch and microwave source.
The amplified microwave signal went through a 50~$\Omega$ PCB and reached the BDD circuits via a $\phi$ = 25 $\mu$m gold wire fixed using Ag paste (4922N, DuPont).
A bias magnetic field of 45 G was applied using a permanent magnet.

For the power-dependent Rabi oscillation measurements, chevron mapping, Ramsey interferometry, and spin relaxation ($T_1$) measurements, the NV centers in BDD samples were optically initialized into the $m_s=0$ state by applying a 50-$\mu$s-width laser pulse with 8-mW power.
The transmitted microwave powers after PCB were monitored using a spectrum analyzer (E4440A, Agilent) while varying the amplitudes of microwave in AWG.
Detailed microwave power values corresponding to each amplitude are presented in Table~\ref{t_odmr}.

\begin{table}[htb!]
\caption{\label{t_odmr} Power and Rabi frequency of microwave amplitude in AWG}
\begin{ruledtabular}
\begin{tabular}{ccc}
AWG Amplitude & Power (mW)& Frequency (MHz)\\
\hline
0.2 & 160.32 & 3.25\\
0.3 & 349.95 & 4.78\\
0.4 & 517.61 & 6.22\\
0.45 & 645.65 & 6.83\\
0.5 & 772.68 & 7.45\\
0.55 & 912.01 & 8.04\\
0.6 & 1054.39 & 8.58\\
0.65 & 1174.90 & 9.04\\
0.7 & 1309.18 & 9.48\\
0.8 & 1520.55 & 10.2\\
0.9 & 1682.67 & 10.6\\
0.99 & 1770.11 & 10.9\\
\end{tabular}
\end{ruledtabular}
\end{table}


\bibliography{NV2}

\end{document}